\documentclass[12pt]{article}
\usepackage{color}
\usepackage{amsmath}
\usepackage{amssymb}
\usepackage{cite}
\catcode`\@=11
\@addtoreset{equation}{section}

\begin{document}
\def\a{\alpha}
\def\b{\beta}
\def\c{\varepsilon}
\def\d{\delta}
\def\e{\epsilon}
\def\f{\phi}
\def\g{\gamma}
\def\h{\theta}
\def\k{\kappa}
\def\l{\lambda}
\def\m{\mu}
\def\n{\nu}
\def\p{\psi}
\def\q{\partial}
\def\r{\rho}
\def\s{\sigma}
\def\t{\tau}
\def\u{\upsilon}
\def\v{\varphi}
\def\w{\omega}
\def\x{\xi}
\def\y{\eta}
\def\z{\zeta}
\def\D{{\mit \Delta}}
\def\G{\Gamma}
\def\H{\Theta}
\def\L{\Lambda}
\def\F{\Phi}
\def\P{\Psi}

\def\S{\Sigma}

\def\o{\over}
\def\beq{\begin{eqnarray}}
\def\eeq{\end{eqnarray}}
\newcommand{\gsim}{ \mathop{}_{\textstyle \sim}^{\textstyle >} }
\newcommand{\lsim}{ \mathop{}_{\textstyle \sim}^{\textstyle <} }
\newcommand{\vev}[1]{ \left\langle {#1} \right\rangle }
\newcommand{\bra}[1]{ \langle {#1} | }
\newcommand{\ket}[1]{ | {#1} \rangle }
\newcommand{\EV}{ {\rm eV} }
\newcommand{\KEV}{ {\rm keV} }
\newcommand{\MEV}{ {\rm MeV} }
\newcommand{\GEV}{ {\rm GeV} }
\newcommand{\TEV}{ {\rm TeV} }
\def\slash#1{\ooalign{\hfil/\hfil\crcr$#1$}}
\def\diag{\mathop{\rm diag}\nolimits}
\def\Spin{\mathop{\rm Spin}}
\def\SO{\mathop{\rm SO}}
\def\O{\mathop{\rm O}}
\def\SU{\mathop{\rm SU}}
\def\U{\mathop{\rm U}}
\def\Sp{\mathop{\rm Sp}}
\def\SL{\mathop{\rm SL}}
\def\tr{\mathop{\rm tr}}

\newcommand\red[1]{{\textcolor{red}{#1}}}

\baselineskip 0.7cm

\begin{titlepage}

\begin{flushright}
\parbox{4.2cm}
{UCB-PTH-09/06 \\
OIQP-09-02 \\
NSF-KITP-09-16}
\end{flushright}

\vskip 1.35cm
\begin{center}
{\large \bf
Non-Relativistic M2-brane Gauge Theory \\
 and New Superconformal Algebra
}
\vskip 1.2cm
Yu Nakayama${}^{1}$, Makoto Sakaguchi${}^{2}$ and Kentaroh Yoshida${}^{3}$

\vskip 0.4cm
${}^{1}${\it Berkeley Center for Theoretical Physics and Department of Physics,
\\
University of California, Berkeley, California 94720-7300, USA
}

${}^2${\it Okayama Institute for Quantum Physics\\
1-9-1 Kyoyama, Okayama 700-0015, Japan}

${}^3${\it Kavli Institute for Theoretical Physics, \\ 
University of California, Santa Barbara, CA 93106, USA}

\vskip 1.5cm

\abstract{We study non-relativistic limits of the $\mathcal{N}=6$
Chern-Simons-Matter theory that arises as a low-energy limit of the
M2-brane gauge theory with background flux. The model admits several
different non-relativistic limits and we find that the maximal
supersymmetry we construct has 14 components of supercharges, which is a
novel example of non-relativistic superconformal algebra in $(1+2)$
dimension. We also investigate the other limits that realize less
supersymmetries.
}
\end{center}
\end{titlepage}

\setcounter{page}{1}
\section{Introduction}

The ubiquity of the Chern-Simons-Matter system has been much appreciated
in recent studies of theoretical physics. On one corner of the
theoretical physics, i.e. in string theory, the M2-brane mini-revolution
\cite{Gustavsson:2007vu,Bagger:2007jr,Bagger:2007vi,Aharony:2008ug}
has created a novel class of gauge-gravity correspondences based on the
Chern-Simons-Matter theory, and we believe that it will eventually bring
us deeper understanding of the M-theory itself. On the other corner of
the theoretical physics, i.e. in condensed matter physics, the
Chern-Simons-Matter theory has been long known to give an indispensable
tool to analyze the effective theory that appears in the quantum Hall
effects.

The natural question that connects these two distinguishing branches of
theoretical physics would be: Can we understand the quantum Hall effect
from M2-brane gauge theory? \footnote{Fractional quantum Hall effect 
has been discussed in \cite{Fujita} by using the edge states in the ABJM 
model \cite{Aharony:2008ug} and other D-brane setups.} 
The question is much like whether we can understand the QCD from the 
string theory. Although it is true that the quantum Hall effect in the effective 
Chern-Simons-Matter system is not supersymmetric (like real QCD) and the 
rank of the gauge group is just Abelian, we expect that qualitative features 
of such a theory can be extracted from the non-relativistic limit of these 
M2-brane gauge theories.

For example, one can use the ``Seiberg duality'' of $\mathcal{N}=2$
Chern-Simons-Matter theory \cite{Aharony:2008gk,Giveon:2008zn} to
translate a level $k$ $U(1)$ Chern-Simons theory with one fundamental
matter multiplet to a level $k$ $U(k)$ Chern-Simons theory with one
fundamental matter multiplet coupled with a singlet supermultiplet. It
may be possible to study the large $k$ behavior from the string theory
because the latter dual theory is strongly coupled from the gauge
theory viewpoint.

However, the real hurdle in this scenario lies in the non-relativistic
limit, which is the main scope of this paper. Even the supersymmetry (SUSY) can
be completely broken in the limiting procedure, depending on the
specific non-relativistic limit that we choose. It is furthermore not
\textit{a priori} obvious how many supersymmetries can be realized in a given
non-relativistic conformal limit. We note that the complete
classification of the non-relativistic superconformal algebra is still
unavailable. Unlike the relativistic superconformal system, there seem a
lot more possibilities. Indeed, we will find many in this paper.

In \cite{Sakaguchi:2008rx,Sakaguchi:2008ku}, they studied the
non-relativistic superconformal algebra embedded in one-dimensional
higher dimensional relativistic superconformal algebra. This is a
standard way to realize the bosonic counterpart: Schr\"odinger algebras
inside a relativistic conformal algebra (or AdS
algebra)\footnote{Non-relativistic conformal algebra
\cite{Hagen:1972pd,Niederer:1972zz,Perroud:1977qh,Barut:1981mt} is
sometimes called Schr\"odinger algebra because it was originally found
as the maximal symmetry of a free Schr\"odinger equation. See also
\cite{Hussin:1986cc,Jackiw:1990mb,Jackiw:1992fg,Henkel:1993sg,Dobrev,
Mehen:1999nd,Nishida:2007pj} for further investigations.}. Some
non-relativistic superconformal algebras we obtain in this paper are not
included in their list. Furthermore, the explicit construction of the
non-relativistic superconformal field theories is non-trivial even when
the algebra is known (see \cite{Leb lanc:1992wu,Nakayama:2008qm,Nakayama:2008qz,NRSY}
for some previous attempts from the field theory).

With this motivation, we study the non-relativistic limit of
$\mathcal{N}=6$ Chern-Simons-Matter theory \cite{Aharony:2008ug} (known
as ABJM model). The model is a candidate dual gauge theory for M2-branes
in orbifold space. We introduce the background 4-form flux that yields
the maximal supersymmetric mass deformation
\cite{Hosomichi:2008jd,Hosomichi:2008jb,Gomis:2008vc}. The
non-relativistic limit of the theory gives a novel supersymmetric
Chern-Simons-Matter theory with maximum $14$ supercharges. We also
obtain less supersymmetric limits, which include the supersymmetric
theory without any superconformal charges (but invariant under the full
bosonic Schr\"odinger group).

The organization of the paper is as follows. In section 2, we study the
maximal supersymmetric mass deformation of the ABJM model. In section 3,
we take the maximal supersymmetric non-relativistic limit of the
mass-deformed ABJM model, and investigate the non-relativistic
superconformal algebra. In section 4, we examine other possible
non-relativistic limits, which yield less supersymmetric theories. 
In section 5, we give some discussions and conclude the paper. In Appendix A,
we have summarized our spinor convention in $(1+2)$ dimension.
We discuss the consistency of the truncation in Appendix B.

\section{Mass deformed ABJM model}

ABJM model describes a low energy effective theory on the M2-branes
probing the $\mathbf{C}_4/\mathbf{Z}_k$ orbifold. It is a $U(N) \times
U(N)$ Chern-Simons quiver gauge theory with bi-fundamental matter
fields. The model has the manifest $\mathcal{N}=6$ superconformal
symmetry (with 24 supercharges).

Our starting point is the relativistic action for the ABJM model given by
\begin{align}
 S&=  \int\! d^3x\,\biggl[\frac{k}{4\pi} \epsilon^{\mu\nu\lambda} 
\mathrm{Tr} ( A_\mu \partial_\nu A_\lambda + \frac{2i}{3}A_\mu A_\nu A_\lambda 
 - \hat{A}_\mu \partial_\nu \hat{A}_\lambda 
-\frac{2i}{3} \hat{A}_\mu \hat{A}_\nu \hat{A}_\lambda ) \cr
& \qquad \qquad \qquad - \mathrm{Tr} D_\mu X_A^\dagger D^\mu X^A 
- i\mathrm{Tr}\bar{\Psi}^{A} \gamma^\mu D_\mu \Psi_A 
- V_{\rm bos} -V_{\rm fer} \biggr] \ ,
\end{align}
where the bosonic potential is given by
\begin{align}
V_{\rm bos} &= -\frac{4\pi^2}{3k^2}\, 
\mathrm{Tr}(X^AX_A^\dagger X^BX_B^\dagger X^CX_C^\dagger 
+X_A^\dagger X^A X_B^\dagger X^B X_C^\dagger X^C \cr
 & \qquad + 4 X^AX_B^\dagger X^C X_A^\dagger X^BX_C^\dagger 
- 6 X^AX_B^\dagger X^B X_A^\dagger X^C X_C^\dagger ) \ ,
\end{align}
while the fermionic potential is given by
\begin{align}
V_{\rm fer} &= -\frac{2\pi i}{k}\, \mathrm{Tr}
\left[ X_A^\dagger X^A \bar{\Psi}^{B}\Psi_B 
+ X^AX_A^\dagger \Psi_B\bar{\Psi}^{B} \right. \cr
& \qquad - 2X^AX_B^\dagger \Psi_A\bar{\Psi}^{B} 
-2 X_A^\dagger X^B \bar{\Psi}^{A}\Psi_B \cr
 & \qquad - \left. \epsilon^{ABCD} X_{A}^\dagger \Psi_B X_C^\dagger \Psi_D 
+ \epsilon_{ABCD} X^A \Psi^{B\dagger}X^C\Psi^{D\dagger} \right]\,. \label{ferp}
\end{align}
The original ABJM model possesses an $SU(4)_R$ symmetry, under which
a field with the
upper index $A$ ($X^A$ and $\Psi^{\dagger A}$) transforms as $4$
and one with the lower index $A$ ($X_A^\dagger$ and $\Psi_A$) transforms as
$\bar{4}$. The gauge group is $U(N) \times U(N)$ and $X^A$ and $\Psi_A$
transform as $(N,\bar{N})$ and $X_A^\dagger$ and $\Psi^{\dagger A}$
transform as $(\bar{N},N)$. The model is parametrized by one integer $k$
given by the level of the Chern-Simons action.

The ABJM action is invariant under the $\mathcal{N}=6$ SUSY
transformation \cite{Gaiotto:2008cg,Terashima:2008sy}
\begin{align}
\delta X^A &= \bar{\epsilon}_i (\Gamma^{i*})^{AB} \Psi_B \ , \qquad 
\delta X^\dagger_A = -\bar{\Psi}^{\dagger B} \epsilon_i \Gamma^i_{AB} \ ,  \cr
\delta \Psi_A &= -i\gamma^\mu \epsilon_i \Gamma^i_{AB} D_\mu X^B \cr 
& \qquad \quad + \frac{2\pi}{k}(-i\epsilon_i \Gamma^i_{AB}(X^CX^\dagger_C X^B
-X^BX^\dagger_CX^C) + 2i\epsilon_i \Gamma_{CD} X^C X_A^\dagger X^D) \ , \cr
\delta \bar{\Psi}^{A} &= 
-iD^\mu X^\dagger_B  \bar{\epsilon}_i \gamma_\mu (\Gamma^{i*})^{AB} \ , \cr 
&\qquad \quad + \frac{2\pi}{k}( i(X^\dagger_B X^C X_C^\dagger 
- X_C^\dagger X^C X_B^\dagger) \bar{\epsilon}_i (\Gamma^{i*})^{AB} 
-2i X_D^\dagger X^AX_C^\dagger \bar{\epsilon}_i (\Gamma^{i*})^{CD})  \ , \cr
\delta{A}_\mu &= 
-\frac{2\pi}{k}(iX^A \bar{\Psi}^B \gamma_\mu \epsilon_i \Gamma^i_{AB} 
+i \bar{\epsilon}_i (\Gamma^{i*})^{AB}\gamma_\mu \Psi_A X^\dagger_B) \ , \cr
\delta\hat{A}_\mu &= 
\frac{2\pi}{k}(i\bar{\Psi}^AX^B\gamma_\mu \epsilon_i \Gamma^i_{AB} 
+i \bar{\epsilon}_i(\Gamma^{i*})^{AB} \gamma_\mu X^\dagger_A \Psi_B) \ ,
\end{align}
where $\epsilon_i$ (for $i=1,\cdots,6$) are six independent Majorana
fermions. We take the explicit form of gamma matrices $\Gamma^i_{AB}$ as
\begin{eqnarray}
& \Gamma^1 = \sigma_2 \otimes 1\,, \qquad 
\Gamma^2 = -i\sigma_2 \otimes \sigma_3\,, \qquad   
\Gamma^3 = i\sigma_2 \otimes \sigma_1\,, \nonumber \\ 
& \Gamma^4 = -\sigma_1 \otimes \sigma_2\,, \qquad  
\Gamma^5 = \sigma_3 \otimes \sigma_2\,, \qquad  
\Gamma^6 = -i 1 \otimes \sigma_2 \,.
\end{eqnarray}
These chiral $SO(6)$ gamma matrices are the intertwiner between the
$SU(4)$ antisymmetric representation (with the reality condition) and the
$SO(6)$ (real) vector representation. Note that
$\frac{1}{2}\epsilon^{ABCD} \Gamma_{CD}^i = -(\Gamma^{i*})^{AB}$. The
model is also invariant under the conformal transformation, so that the
theory has 12 additional superconformal charges \cite{Bandres:2008ry}.

The mass deformation of the ABJM model was studied in
\cite{Hosomichi:2008jd,Hosomichi:2008jb,Gomis:2008vc}. We focus on the
maximally supersymmetric mass deformation,
\begin{align}
V_{\rm mass} &= m^2 \mathrm{Tr} (X^{\dagger}_ A X^A) 
+ im \mathrm{Tr}(\bar{\Psi}^{a} \Psi_a - \bar{\Psi}^{a'} \Psi_{a'}) \cr 
& \qquad - \frac{4\pi}{k}m \mathrm{Tr} (X^aX_{[a}^\dagger X^bX^\dagger_{b]} 
- X^{a'}X^\dagger_{[a'}X^{b'}X^\dagger_{b']}) \ , \label{mass}
\end{align}
which  breaks the $SU(4)$
R-symmetry down to the $SU(2)\times SU(2)\times U(1)$.
We set $A =
(a,a')$, where $a$ and $a'$ are two $SU(2)$ indices,
and we have introduced the following notation 
\[
X_{[a}X_{b]} \equiv X_a X_b-X_b X_a \ .
\]

Though the mass term breaks the $SU(4)$ R-symmetry down to the
$SU(2)\times SU(2) \times U(1)$, the $\mathcal{N}=6$ SUSY
remains once we add the SUSY transformation
\begin{eqnarray}
&& \delta_m \Psi_a = im \epsilon_i \Gamma^{i}_{aB} X^B \ , \qquad ~~~~
\delta_m \bar{\Psi}^a =-im \bar{\epsilon}_i (\Gamma^{i*})^{aB}
X^\dagger_B  \ , 
\nonumber \\ 
&& \delta_m \Psi_{a'} = -im \epsilon_i \Gamma^{i}_{a'B} X^B \ , \qquad 
\delta_m \bar{\Psi}^{a'} = im \bar{\epsilon}_i (\Gamma^{i*})^{a'B} 
X^\dagger_B \ .
\end{eqnarray}
The mass deformation obviously breaks the conformal invariance, so the
$12$ superconformal generators are lost accordingly.  From the M-theory
viewpoint, the mass deformation corresponds to turning on a background
4-form flux in the bulk. After taking the mass deformation, the theory
has multiple vacua including the broken (Higgs) phase, but we focus on
the unbroken phase in the following non-relativistic limit analysis.

\section{Non-relativistic limit}

There are several possible ways to take a non-relativistic limit of the
relativistic action. We first investigate the
non-relativistic limit which preserves the maximal SUSY.
It turns out that the
non-relativistic limit preserves $14$ supercharges (including 2
superconformal charges).

\subsection{Action}

We begin with the bosonic part. The relativistic scalar field $X^A$ can
be decomposed into two non-relativistic scalar fields $\phi^A$ and
$\hat{\phi}^{*A}$ \cite{Jackiw:1990mb,Leblanc:1992wu} as
\begin{eqnarray}
X^A = \frac{1}{\sqrt{2m}} \left( e^{-imt} \phi^A 
+ e^{imt} \hat{\phi}^{*A}\right) \ , \label{bosan}
\end{eqnarray}
where $\phi^A$ describes a particle degree of freedom and $\hat{\phi}_A$
describes an anti-particle degree of freedom. To obtain the maximal
 SUSY transformation, we discard $\hat{\phi}_A$ and only keep
$\phi^A$.\footnote{In the later section, we will investigate other
choices of the non-relativistic limit to obtain less supersymmetric
theories. We discuss the consistency of the truncation in Appendix B.}
After the substitution of our ansatz \eqref{bosan}, the kinetic part of
the original relativistic action is replaced by the Schr\"odinger
action:
\begin{eqnarray}
i\mathrm{Tr} (\phi^\dagger_A D_0 \phi^A) 
- \frac{1}{2m} \mathrm{Tr} (D_i\phi^\dagger_A D_i \phi^A) \ . 
\end{eqnarray}

Similarly, the relativistic fermion field $\Psi^A$ can be decomposed
into non-relativistic two-component spinor fields $\psi_{\alpha A}$
and $\hat{\psi}^{A}_{\alpha}$ in the following form:
\begin{eqnarray}
\Psi_A = e^{-imt} \psi_A + e^{imt} \sigma_2 \hat{\psi}^{*}_A  \ .
\end{eqnarray}
Again, in order to obtain the maximal SUSY theory, we discard
the anti-particle degrees of freedom $\hat{\psi}_A$. Actually, only the
half of the spinor components are dynamical in the non-relativistic
limit. To see this, we note that the Dirac equation
\begin{eqnarray}
\left( 
  \begin{array}{cc}
   iD_0 + m & D_+ \\ 
   D_-  & -iD_0 + m
  \end{array}
\right) 
\left( 
  \begin{array}{c}
   \Psi_{1a} \\ 
    \Psi_{2a}  
  \end{array}
\right) = 0    \qquad (D_{\pm} \equiv D_1  \pm i D_2) \ , 
\end{eqnarray}
is decomposed into the two equations:
\begin{align}
2m \psi_{1a} + D_+ \psi_{2a} = 0 \ , \qquad 
D_- \psi_{1a} - i D_0 \psi_{2a} = 0 \ , 
\end{align}
in the non-relativistic limit. We can replace the first component of the
non-relativistic spinor $\psi_{1a}$ by $-\frac{D_+}{2m}
\psi_{2a}$. Then, the non-relativistic equation for the second component
of the fermion is given by the Pauli equation:
\begin{eqnarray}
iD_0 \psi_{2a} = -\frac{D_-D_+}{2m} \psi_{2a} \ .
\end{eqnarray}

In the same way, the Dirac equation for $\Psi_{a'}$ is given by
\begin{eqnarray}
\left( 
  \begin{array}{cc}
   iD_0 - m & D_+ \\ 
   D_-  & -iD_0 - m
  \end{array}
\right) 
\left( 
  \begin{array}{c}
   \Psi_{1a'} \\ 
    \Psi_{2a'}  
  \end{array}
\right) = 0 \ ,
\end{eqnarray}
and in the non-relativistic limit, it becomes
\begin{align}
iD_0 \psi_{1a'} + D_+ \psi_{2a'} = 0 \ , \qquad  
D_- \psi_{1a'} -2m \psi_{2a'} = 0 \ .
\end{align}
We can replace the second component of the non-relativistic spinor
$\psi_{2a'}$ by $\frac{D_-}{2m} \psi_{1a'}$, and the first equation
yields the Pauli equation:
\begin{eqnarray}
iD_0 \psi_{1a'} = -\frac{D_+D_-}{2m} \psi_{1a'} \ .
\end{eqnarray} 
In the following, we drop the subscript $1$ (for $\psi_{a'}$) and $2$
(for $\psi_a$) with the above substitution implicitly assumed.

We now present the non-relativistic ABJM action obtained by substituting
the above non-relativistic ansatz. We only keep the quartic potential
terms and neglect the sextic terms that are irrelevant deformations in
the non-relativistic superconformal limit
\cite{Jackiw:1990mb}\cite{Leblanc:1992wu}.

Due to the topological nature, there is no change in the
Chern-Simons term:
\begin{eqnarray}
S_{\rm CS} = \frac{k}{4\pi} \int\! dt d^2x\, \epsilon^{\mu\nu\lambda}\, 
\mathrm{Tr} \left[ A_\mu \partial_\nu A_\lambda 
+ \frac{2i}{3}A_\mu A_\nu A_\lambda 
 - \hat{A}_\mu \partial_\nu \hat{A}_\lambda 
-\frac{2i}{3} \hat{A}_\mu \hat{A}_\nu \hat{A}_\lambda \right] \ . \label{CSa}
\end{eqnarray}
The kinetic terms for bosons and fermions are given by
\begin{align}
S_{\rm kin} &= \int\! dt d^2x\, \left[i\mathrm{Tr} 
(\phi^\dagger_A D_0 \phi^A) 
- \frac{1}{2m} \mathrm{Tr} (D_i\phi^\dagger_A D_i \phi^A) \right. \cr
& \qquad \quad + \left. i\mathrm{Tr} (\psi^{\dagger A} D_0 \psi_A) 
+ \frac{1}{2m} \mathrm{Tr} (\psi^{\dagger a} D_- D_+ \psi_a 
+ \psi^{\dagger a'} D_+D_- \psi_{a'}) \right] . \label{kine}
\end{align}
We can also rewrite the Pauli terms as 
\begin{align}
&  \frac{1}{2m} \mathrm{Tr} (\psi^{\dagger a} D_- D_+ \psi_a 
+ \psi^{\dagger a'} D_+D_- \psi_{a'}) \cr
=& \frac{1}{2m} \mathrm{Tr}\left[
\psi^{\dagger a} {
\left(D^2_i\psi_a - F_{12}\psi_a + \psi_a\hat{F}_{12} \right)
} 
\right] 
+ \frac{1}{2m} \mathrm{Tr}\left[\psi^{\dagger a'} {
\left(D^2_i\psi_{a'}  + F_{12}\psi_{a'}  -\psi_{a'}  \hat{F}_{12} \right)
}
\right] \ . \notag
\end{align}
The non-relativistic fields $\phi^a$, $\phi^{a'}$, $\psi_a$ and
$\psi_{a'}$ all transform as $(N,\bar{N})$ under $U(N) \times U(N)$.

Let us move on to the potential part. As we have mentioned, we discard
the irrelevant sextic potential and we only keep the marginal quartic
terms.\footnote{Note the classical scaling dimension of the
non-relativistic fields $D(\phi^a) = D(\phi^{a'})= D(\psi_a) =D(\psi_{a'})=1$.} The bosonic
potential comes from the supersymmetric completion of the mass term in
\eqref{mass}, leading to
\begin{eqnarray}
S_{\rm bos} = \frac{\pi}{km} \int\! dt d^2x\,  
\mathrm{Tr} 
\left(\phi^a\phi^\dagger_{[a} \phi^b\phi^\dagger_{b]} 
- \phi^{a'}\phi^\dagger_{[a'} \phi^{b'} \phi_{b']}^\dagger\right) \ . 
\label{bosa}
\end{eqnarray}

The fermionic potential comes from the non-relativistic limit of \eqref{ferp}: 
\begin{align}
S_{\rm fer} = \frac{\pi }{km}&\int\! dt d^2x\,  \mathrm{Tr} \left[
(\phi^\dagger_a\phi^a + \phi^\dagger_{a'} \phi^{a'})(\psi^{\dagger
b}\psi_b - \psi^{\dagger b'} \psi_{b'}) \right. \nonumber \\ 
& +(\phi^a\phi^\dagger_a
+\phi^{a'}\phi^\dagger_{a'} )(\psi_b \psi^{\dagger b} - \psi_{b'}
\psi^{\dagger b'} )  \nonumber \\
&  - 2 \phi^a\phi^\dagger_b
\psi_a\psi^{\dagger b} + 2 \phi^{a'} \phi^\dagger_{b'} \psi_{a'}
\psi^{\dagger b'} 
- 2 \phi^\dagger_a \phi^b\psi^{\dagger a}\psi_b +2
\phi^\dagger_{a'} \phi^{b'} \psi^{\dagger a'} \psi_{b'} \nonumber \\ 
&  
-i\epsilon^{ab}\epsilon^{c'd'} \phi^\dagger_a \psi_b \phi^\dagger_{c'}
\psi_{d'} 
- i \epsilon^{bc}\epsilon^{a'd'}\phi^\dagger_{a'} \psi_b
\phi^\dagger_c\psi_{d'} \nonumber \\ 
& +i\epsilon^{a'b'}\epsilon^{cd} \phi^\dagger_{a'}
\psi_{b'} \phi^\dagger_{c} \psi_{d}+
i\epsilon^{b'c'}\epsilon^{ad}\phi^\dagger_{a}
\psi_{b'}\phi^\dagger_{c'}\psi_{d} \nonumber \\ 
&  
+i\epsilon_{ab}\epsilon_{c'd'} \phi^a \psi^{\dagger b} \phi^{c'}
\psi^{\dagger d'} + i \epsilon_{bc}\epsilon_{a'd'}\phi^{a'}
\psi^{\dagger b} \phi^c\psi^{\dagger d'} \nonumber \\ 
& \left. -i\epsilon_{a'b'}\epsilon_{cd}
\phi^{a'} \psi^{\dagger b'} \phi^{c} \psi^{\dagger d}-
i\epsilon_{b'c'}\epsilon_{ad}\phi^{a} \psi^{\dagger
b'}\phi^{c'}\psi^{\dagger d} \right] \ . \label{fera}
\end{align}
Here, we have dropped the higher dimensional terms including the
derivatives of fermions. The final non-relativistic ABJM action is given
by the sum of \eqref{CSa}, \eqref{kine}, \eqref{bosa} and \eqref{fera}.

\subsection{Bosonic symmetry}

Let us investigate the symmetry of the non-relativistic ABJM
model. First of all, the model is invariant under the bosonic
Schr\"odinger symmetry ($+$ some internal symmetries):
\begin{itemize}

\item time translation: $\delta t = -a$
\begin{align}
\delta \phi^A &= a {D_0 \phi^A} \ , \qquad 
\delta \psi_A = a{D_0} \psi_A \ , \cr 
\delta A_0  &= \delta \hat{A}_0 = 0 \ , \qquad \delta A_i = a F_{0i} \ , 
\qquad 
\delta \hat{A}_i = a \hat{F}_{0i} \ ,
\end{align}
with the conserved charge (Hamiltonian)
\begin{align}
H &= \int\! d^2x\, \left[ \frac{1}{2m} \mathrm{Tr}(D_i \phi^\dagger_A
 D_i\phi^A + D_i \psi^{\dagger A} D_i \psi_A) 
+ \frac{1}{2m}\mathrm{Tr}(\psi^{\dagger a}{(F_{12}\psi_a 
- \psi_a\hat{F}_{12})})  \right. \cr
& \qquad\qquad\quad 
- \left.  \frac{1}{2m} \mathrm{Tr}\left(\psi^{\dagger a'} 
{(F_{12}\psi_{a'} - \psi_{a'}\hat{F}_{12})}
\right) + V_{\rm bos} + V_{\rm fer} \right] \ .  \label{hamil}
\end{align}

 \item spatial translation: $\delta x^i = a^i$  
\begin{align}
\delta \phi^A &=  -a^i D_i {\phi^A} \ , \ \quad 
\delta \psi_A =  -a^iD_i \psi_A \ , \cr 
\delta A_0  &= a^i F_{0i} \ , \quad
\delta \hat{A}_0 =a^i\hat{F}_{0i} \ , \quad  
\delta A_i = \epsilon_{ij}a^j F_{12} \ , \quad  
\delta \hat{A}_i = \epsilon_{ij} a^j \hat{F}_{12} \ ,
\end{align}
with the conserved charge (momentum) 
\begin{align}
P_i &= \int\! d^2x\, p_i~,
~~~
p_i= -\frac{i}{2} 
\mathrm{Tr} \left[ \phi^\dagger_A D_i\phi^A 
- D_i \phi^\dagger_A \phi^A + \psi^{\dagger A} D_i \psi_A 
-D_i \psi^{\dagger A} \psi_A \right] \ .
\end{align}

\item infinitesimal rotation: $\delta x_i = -\theta \epsilon_{ij} x^j$
\begin{align}
& \hspace*{-0.2cm}
\delta \phi^A = \theta \epsilon_{ij} x^i D^j \phi^A \ , ~~
\delta \psi_a = \theta \epsilon_{ij} x^i D^j \psi_a 
{+} 
\frac{i}{2} \theta \psi_a \ , ~~ 
\delta \psi_{a'} = \theta \epsilon_{ij} x^i D^j \psi_{a'} 
{-}
 \frac{i}{2} \theta \psi_{a'} \ , \cr
& \hspace*{-0.2cm} \delta A_0 = -\theta \epsilon_{ij} x^i F_{0j} \ , \quad  
\delta\hat{A}_0  = -\theta \epsilon_{ij} x^i \hat{F}_{0j} \ ,  \quad 
\delta A_{i} = -\theta x_i {F_{12}} \ , \quad  
\delta \hat{A}_{i} = -\theta x_i {\hat{F}_{12}} \ , 
\end{align}
with the conserved charge ($U(1)$ angular momentum)\footnote{We have
added the separately conserved $U(1)_F$ to obtain the conventional spin $1/2$
of fermions. Actually, in two spatial-dimension, the addition of
arbitrary amount of $U(1)_F$ (or $U(1)_B$) does not change the
Schr\"odinger algebra.}
\begin{eqnarray}
J = -\int\! d^2x\, \left[ \epsilon^{ij} x_i p_j 
{+}
 \frac{1}{2} \mathrm{Tr} (\psi^{\dagger a} 
\psi_a - \psi^{\dagger a'} \psi_{a'}) \right] \ .
\end{eqnarray}
\item total number density (actually a part of the gauge symmetry)
\begin{align}
\delta \phi^A = -i\alpha m \phi^A \ , \quad \delta 
\psi_A = -i\alpha m \psi_A \ , \quad 
\delta{A}_\mu = \delta\hat{A}_\mu = 0 \ ,
\end{align}
with the conserved charge (total mass operator)
\begin{eqnarray}
M = m\int\! d^2x \, \rho \ , \quad 
\rho = \mathrm{Tr}(\phi_A^\dagger \phi^A + \psi^{\dagger A} \psi_A) \ . 
\end{eqnarray}
\item infinitesimal Galilean boost: $\delta x^i = -v^it$
\begin{align}
& \delta \phi^A = -(im v^i x_i -tv^iD_i)\phi^A \ , \quad 
\delta \psi_A =-(im v^i x_i -t v^iD_i)\psi_A\,, \cr 
& \delta{A}_0 = -t v^i F_{0i} \ , \quad \delta\hat{A}_0 = -t v^i
 \hat{F}_{0i}\,,  
\cr  
& \delta{A}_i = -t \epsilon_{ij} v^j F_{12} \ , \quad  
\delta\hat{A}_i = -t \epsilon_{ij} v^j \hat{F}_{12} \ , 
\end{align}
with the conserved charge 
\begin{eqnarray}
G_i = \int\! d^2x\, \left[ -t p_i + m x_i \rho\right] \ . 
\end{eqnarray}
\item infinitesimal dilatation: $\delta t = 2{\alpha}t$, $\delta x^i =
      {\alpha} x^i$ 
\begin{align}
& \delta \phi^A = -\alpha (1+ x^iD_i + 2t{D_0}) \phi^A \ , \quad  
\delta \psi_A = -\alpha(1+x^iD_i + 2t {D_0}) \psi_A \ , \cr 
& \delta A_0 = \alpha x^i F_{0i} \ , \quad 
\delta \hat{A}_0  =  \alpha x^i \hat{F}_{0i} \ , \cr
& \delta{A}_i = \alpha (\epsilon_{ij} x^j F_{12} - 2 t F_{0i}) \ , \quad  
\delta\hat{A}_i = \alpha(\epsilon_{ij} x^j \hat{F}_{12} - 2 t \hat{F}_{0i}) \ ,
\end{align}
with the conserved charge 
\begin{eqnarray}
D = -2t H +  \int\! d^2x\, x^i p_i \ . 
\end{eqnarray}
\item infinitesimal special conformal transformation: $\delta t = -at^2$,
      $\delta x^i = -at x^i$
\begin{align}
& \delta \phi^A = \left(at -\frac{i}{2} m a x^2 
+ at x^i D_i + at^2 {D_0}\right) \phi^A \cr
& \delta \psi_A = \left(at -\frac{i}{2} ma x^2 
+ at x^i D_i + at^2 {D_0} \right) \psi_A \cr
& \delta A_0 = -at x^i F_{0i} \ , \quad 
\delta \hat{A}_0 =- at x^i \hat{F}_{0i} \cr
& \delta A_i =- a t \epsilon_{ij} x^j F_{{12}} + at^2 F_{0i} \ , \quad  
\delta \hat{A}_i = -at \epsilon_{ij} x^j \hat{F}_{{12}} 
+ a t^2 \hat{F}_{0i} \ , 
\end{align}
with the conserved charge
\begin{eqnarray}
K = -t^2 H - t D +\frac{1}{2} m \int\! d^2x\, x^2 \rho \ .
\end{eqnarray}
\end{itemize}

These generators satisfy the Schr\"odinger algebra\footnote{The Poisson
bracket (more precisely Dirac bracket) is defined by $[F,G]_{PB} =
-i\left(-\frac{\delta F}{\delta \phi^*} \frac{\delta G}{\delta \phi} +
\frac{\delta F}{\delta \phi} \frac{\delta G}{\delta \phi^*}\right) -i
\left(\frac{\delta^r F}{\delta \psi^*} \frac{\delta^l G}{\delta \psi} +
\frac{\delta^r F}{\delta \psi} \frac{\delta^l G}{\delta \psi^*}\right)$,
where $\frac{\delta^r}{\delta \psi}$ denotes the right derivative and
$\frac{\delta^l}{\delta \psi}$ denotes the left derivative. We further
replace the Poisson bracket with the quantum mechanical
(anti-)commutator $[F,G]_{PB} \to -i[F,G]$ or $-i\{F,G\}$.}
\begin{eqnarray}
&&
i[J,P_i]=\epsilon_{ij}P_j~,~~~
i[J,G_i]=\epsilon_{ij}G_j~,~~~
i[P_i,G_j]=\delta_{ij}M~,~~~
i[H,G_i]=P_i~,~~~
\cr&&
i[D,H]=-2H~,~~~
i[D,K]=2K~,~~~
i[H,K]=D~,~~~
i[K,P_i]=-G_i~,~~~
\cr&&
i[D,G_i]=G_i~,~~~
i[D,P_i]=-P_i~.
\end{eqnarray}
To derive these, as in \cite{Jackiw:1990mb}, it is useful to note that
$A_i$ and $\hat A_i$ are solved by $A_+=\hat A_+=0$,
$A_-=-\frac{4\pi}{k} i \partial_{-}
\int d^2y G(x-y) (\phi^A\phi^\dagger_A-\psi_A\psi^{\dagger A})(y)$
and $\hat A_-=-\frac{4\pi}{k} i \partial_{-}
\int d^2y G(x-y) (\phi_A^\dagger \phi^A+\psi^{\dagger A}\psi_A)(y)$
where $G(x-y)=\frac{1}{2\pi}\log|x-y|$.

In addition, the model possesses some internal global symmetries:
\begin{itemize} 

\item $U(1)_B \times U(1)_F$
\begin{align}
& \delta \phi^a = -i\alpha \phi^a \ , \quad 
\delta \phi^{a'} = i\alpha \phi^{a'} \ , \quad 
\delta \psi_a = -i\beta \psi_a \ , \quad  
\delta \psi_{a'} = i\beta \psi_{a'} \ , \cr
& \delta A_\mu = \delta \hat{A}_\mu = 0 \ ,
\end{align}
with the conserved charges
\begin{align} 
Q_B = \int\! d^2x\, \mathrm{Tr} 
(\phi^\dagger_a \phi^a - \phi^\dagger_{a'} \phi^{a'}) \ , \qquad 
Q_F = \int\! d^2x\, \mathrm{Tr} 
(\psi^{\dagger a} \psi_a - \psi^{\dagger a'} \psi_{a'}) \ .
\end{align}
We have used $Q_F$ to improve the $U(1)$ angular momentum. The diagonal
part $\alpha = \beta$ is a part of the gauge symmetry.

\item $SU(2)\times SU(2)$ R-symmetry

The first $SU(2)$ is generated by
\begin{align}
\delta \phi^a = i\alpha^i(\sigma_i)^{a}_{\ b} \phi^b \ , \quad 
\delta \psi_a = -i\alpha^{i}(\sigma_i^*)_{a}^{\ b} \psi_b \ , \quad 
\delta (\mathrm{others}) = 0 \ .
\end{align} 
The corresponding generator is 
\begin{align}
R^{(1)}_i &= \int\! d^2x\, \mathrm{Tr} \left( 
\phi^\dagger_a (\sigma_i)^{a}_{\ b} \phi^b 
- \psi^{\dagger a} (\sigma_i^*)_{a}^{\ b} \psi_b \right) \ . 
\end{align}
Similarly, 
\begin{align}
\delta \phi^{a'} = i\alpha^i(\sigma_i)^{a'}_{\ b'} \phi^{b'} \ , \quad  
\delta \psi_{a'} = -i\alpha^{i}(\sigma_i^*)_{a'}^{\ b'} \psi_{b'} \ , \quad 
\delta (\mathrm{others}) = 0 \ .
\end{align} 
The corresponding generator is 
\begin{align}
R^{(2)}_i &= \int\! d^2x\, \mathrm{Tr} \left( 
\phi^\dagger_{a'} (\sigma_i)^{a'}_{\ b'} \phi^{b'} 
- \psi^{\dagger a'} (\sigma_i^*)_{a'}^{\ b'} \psi_{b'} \right) \ .  
\label{su2r}
\end{align}
The above global internal symmetries commute with all the bosonic generators
of the Schr\"odinger algebra.

\end{itemize}

\subsection{Supersymmetry}

The non-relativistic limit of the mass deformed ABJM model has the non-relativistic
supersymmetry induced from the supersymmetry of the original
relativistic theory.  Let us first begin with the kinematical SUSY. The first order supersymmetry is obtained by the direct
non-relativistic limit of the relativistic supersymmetry. They are
generated by the following charges
\begin{align}
\epsilon \mathcal{Q}
 &=i\sqrt{2m}\left[(\epsilon_1+i\epsilon_2)\mathrm{Tr} 
( i\phi^\dagger_1 \psi_2 -i \phi^\dagger_2 \psi_1
 -\phi^{1'}\psi^{\dagger 2'} 
+ \phi^{2'} \psi^{\dagger 1'}) \right. \cr
 &\qquad \quad 
+ \epsilon_3 \mathrm{Tr}(-i\phi^1\psi^{\dagger 2'} +i \phi^2
 \psi^{\dagger 1'} + \phi^\dagger_{1'} \psi_2 -\phi^\dagger_{2'} \psi_1
 ) \cr 
& \qquad \quad  + \epsilon_4 \mathrm{Tr} (\phi^1\psi^{\dagger 2'} +
 \phi^2\psi^{\dagger 1'} 
+i \phi^\dagger_{1'} \psi_2 +i \phi^\dagger_{2'} \psi_1) \cr
& \qquad \quad + \epsilon_5 \mathrm{Tr}(-\phi^1 \psi^{\dagger 1'}+ \phi^2
 \psi^{\dagger 2'} -i \phi^\dagger_{1'} \psi_1 +i \phi^\dagger_{2'}
 \psi_2) \cr 
& \qquad \quad \left. 
+ \epsilon_6(i\phi^1 \psi^{\dagger 1'} +i \phi^2 \psi^{\dagger 2'} 
+\phi_{1'}^\dagger \psi_1 + \phi_{2'}^\dagger \psi_2) \right] \ ,
\end{align}
and similarly by $\epsilon^* \mathcal{Q}^*$ by just complex conjugation.
There are total five independent complex supercharges,\footnote{Note
that $\epsilon_1 - i\epsilon_2$ does not appear in the first
supercharges, which results in the emergence of the second dynamical SUSY.} and we relabel them so that
\begin{eqnarray}
Q^{\hat{i}}_1 \equiv \sqrt{2m} \int\! d^2x\, j_{\hat{i}} \qquad  
(\hat{i} = 0,3 \cdots, 6) \ ,
\end{eqnarray}
where
\begin{align}
j_0 &= \mathrm{Tr}(i\phi^\dagger_1 \psi_2 -i \phi^\dagger_2 \psi_1 
- \phi^{1'}\psi^{\dagger 2'} + \phi^{2'} \psi^{\dagger 1'}) \ , \cr
j_3 &=\mathrm{Tr}(-i\phi^1\psi^{\dagger 2'} +i \phi^2 \psi^{\dagger 1'} 
+ \phi^\dagger_{1'} \psi_2 - \phi^\dagger_{2'} \psi_1) \ , \cr
j_4 &=\mathrm{Tr} (\phi^1\psi^{\dagger 2'}+ \phi^2\psi^{\dagger 1'} 
+i \phi^\dagger_{1'} \psi_2 +i \phi^\dagger_{2'} \psi_1) \ , \cr
j_5 &= \mathrm{Tr} (-\phi^1 \psi^{\dagger 1'} +\phi^2 \psi^{\dagger 2'} 
-i\phi^\dagger_{1'} \psi_1 +i \phi^\dagger_{2'} \psi_2) \ , \cr
j_6 &= \mathrm{Tr}(i\phi^1 \psi^{\dagger 1'} +i  \phi^2 \psi^{\dagger
 2'} 
+ \phi_{1'}^\dagger \psi_1 + \phi_{2'}^\dagger \psi_2) \ .
\end{align}
$Q_1^{0}$ is singlet under the $SU(2) \times SU(2)$ R-symmetry while
$Q_1^i$ $(i=3,\cdots 6)$ transform as $2\times 2$ representations under
the $SU(2) \times SU(2)$.

We can compute the anti-commutation relations as 
\begin{eqnarray}
&& \{Q^0_1 , Q_1^{0*}\} = 2  M \ , \qquad 
\{Q_1^{m *}, Q_1^{n} \} = 2 M\delta^{mn} - 2im R^{mn} \ , \nonumber \\
&& \{Q_1^0, Q_1^{m*} \} = \{Q_1^{\hat{i}}, Q_1^{\hat{j}} \} = 0 \ ,  \nonumber \\ 
&& i[J,Q_1^0] = \frac{i}{2} Q_1^0 \ , \qquad  
i[J, Q_1^m] = \frac{i}{2} Q_1^m \ , \nonumber \\ 
&& { [H,Q_1^{\hat{i}}] =} [P_i,Q_1^{\hat{i}}] = [G_i,Q_1^{\hat{i}}] 
= [D,Q_1^{\hat{i}}] = [K,Q_1^{\hat{i}}] = [M,Q_1^{\hat{i}}] = 0 \ . 
\end{eqnarray}
$R^{mn}$ are particular combinations of the $SU(2)\times SU(2)$
R-charges introduced in \eqref{su2r}:
\begin{align}
R^{34} &= \int d^2 x \mathrm{Tr}(- \psi^{\dagger 2'} \psi_{2'} + \psi^{\dagger 1'}\psi_{1'} + \psi_2 \psi^{\dagger 2} - \psi_1 \psi^{\dagger 1} - \phi^1 \phi_1^\dagger + \phi^2 \phi_2^\dagger - \phi_{1'}^\dagger \phi^{1'} + \phi_{2'}^\dagger \phi^{2'})\,, \cr
R^{35} &= \int d^2 x \mathrm{Tr}(\psi^{\dagger 2'} \psi_{1'} + \psi^{\dagger 1'}\psi_{2'} - \psi_2 \psi^{\dagger 1} - \psi_1 \psi^{\dagger 2} - \phi^1 \phi_2^\dagger - \phi^2 \phi_1^\dagger - \phi_{1'}^\dagger \phi^{2'}- \phi_{2'}^\dagger \phi^{1'}) \,,\cr
 R^{36} &= \int d^2 x \mathrm{Tr}(i \psi^{\dagger 2'} \psi_{1'} - i\psi^{\dagger 1'}\psi_{2'} +i \psi_2 \psi^{\dagger 1} -i \psi_1 \psi^{\dagger 2} +i \phi^1 \phi_2^\dagger -i \phi^2 \phi_1^\dagger -i \phi_{1'}^\dagger \phi^{2'} +i \phi_{2'}^\dagger \phi^{1'})\,, \cr
 R^{45} &= \int d^2 x \mathrm{Tr}(i \psi^{\dagger 2'} \psi_{1'} - i\psi^{\dagger 1'}\psi_{2'} -i \psi_2 \psi^{\dagger 1} +i \psi_1 \psi^{\dagger 2} -i \phi^1 \phi_2^\dagger +i \phi^2 \phi_1^\dagger -i \phi_{1'}^\dagger \phi^{2'} +i \phi_{2'}^\dagger \phi^{1'})\,, \cr
R^{46} &= \int d^2 x \mathrm{Tr}(-\psi^{\dagger 2'} \psi_{1'} - \psi^{\dagger 1'}\psi_{2'} - \psi_2 \psi^{\dagger 1} - \psi_1 \psi^{\dagger 2} - \phi^1 \phi_2^\dagger - \phi^2 \phi_1^\dagger + \phi_{1'}^\dagger \phi^{2'} + \phi_{2'}^\dagger \phi^{1'}) \,,\cr
R^{56} &= \int d^2 x \mathrm{Tr}( -\psi^{\dagger 2'} \psi_{2'} + \psi^{\dagger 1'}\psi_{1'} - \psi_2 \psi^{\dagger 2} + \psi_1 \psi^{\dagger 1} 
{+} \phi^1 \phi_1^\dagger 
{-} \phi^2 \phi_2^\dagger 
{-} \phi_{1'}^\dagger \phi^{1'} 
{+} \phi_{2'}^\dagger \phi^{2'}) \,.
\end{align}

Since the particular combination of the SUSY parameter $\epsilon_1 +
i\epsilon_3$ does not generate the first order kinematical SUSY
transformation, one can construct the second dynamical SUSY
transformation \cite{Leblanc:1992wu,Nakayama:2008qz}. The second SUSY is generated by the supercharge
\begin{eqnarray}
Q_2 =  \frac{1}{\sqrt{2m}} \int\! d^2x\, 
\mathrm{Tr}\left(\phi^{\dagger}_1 D_+\psi_{2} -\phi^{\dagger}_2 D_+\psi_{1} 
-i \phi^{1'} D_+ \psi^{\dagger 2'} 
+i \phi^{2'} D_+\psi^{\dagger 1'} \right) \ .
\end{eqnarray}
The supercharge $Q_2$ is invariant under $SU(2)\times SU(2)$ R-symmetry.
The anti-commutation relations for $Q_2$ can be computed as
\begin{eqnarray}
&& \{Q_2,Q_2^{*}\}=H \ , \quad  
\{Q_1^0,Q_2^*\}= P_- \ , \quad \{Q_1^0,Q_2\}=\{Q_1^m,Q_2^*\}=\{Q_1^m,Q_2\}=0 \ , 
\nonumber \\
&& [P_i, Q_2] = [H,Q_2]=0 \ ,  \quad i[J,Q_2]=-\frac{i}{2}Q_2 \ ,   \nonumber \\
&& i[G_{+},Q_2^*] =-Q_1^{0*} \ , \quad 
i[G_{-},Q_2]=-Q_1^{0} \ , \quad i[D,Q_2] = -Q_2 \ , \nonumber \\
&& i[K,Q_2] =t Q_2 -\sqrt{\frac{m}{2}}\int \!d^2 x\, x^+ j_0 \ , \quad  
[M,Q_2]= [R^{mn},Q_2]=0 \ . \label{anti}
\end{eqnarray}
As expected from the first anti-cummutation relation in \eqref{anti},
we can rewrite the Hamiltonian \eqref{hamil} by using the Gauss law constraints
\begin{eqnarray}
F_{12}=\frac{2\pi}{k}\left(
\phi^A\phi_A^\dag-\psi_A\psi^{\dag A}
\right)~,~~~
\hat F_{12}=\frac{2\pi}{k}\left(
\phi_A^\dag\phi^A+\psi^{\dag A}\psi_A
\right)~,
\end{eqnarray}
in a manifestly semi-positeve definite form:
\begin{eqnarray}
H&=&
\int d^2x  \bigg[
\frac{1}{2m}\mathrm{Tr}\big((D_-\phi^a)^\dag D_-\phi^a
+(D_+\phi^{a'})^\dag D_+\phi^{a'}\big)
\cr&&\qquad\quad
+\frac{1}{2m}\mathrm{Tr}\big((D_+\psi_a)^\dag D_+\psi_a
+(D_-\psi_{a'})^\dag D_-\psi_{a'}
\big)
\cr&&\qquad\quad
+\frac{2\pi}{mk}\mathrm{Tr}\big(
(\epsilon_{ab}\phi^a\psi^{\dag b}
 -i\epsilon^{a'b'}\psi_{b'}\phi^\dag_{a'})^\dag
 (\epsilon_{ab}\phi^a\psi^{\dag b}
 -i\epsilon^{a'b'}\psi_{b'}\phi^\dag_{a'})\big)
\cr&&\qquad\quad
+\frac{2\pi}{mk}\mathrm{Tr}\big(
(\epsilon^{ab}\phi_a^\dag\psi_b+i\epsilon_{a'b'}\psi^{\dag b'}\phi^{a'})^\dag
(\epsilon^{ab}\phi_a^\dag\psi_b+i\epsilon_{a'b'}\psi^{\dag b'}\phi^{a'})
\big)
\bigg]~.
\end{eqnarray}

The commutator of $K$ and $Q_2$ defines the superconformal charge
\begin{eqnarray}
i[K,Q_2]= S \ ,
\end{eqnarray}
so that
\begin{eqnarray}
S=t Q_2-\sqrt{\frac{m}{2}}
\int\! d^2x\, x^+ j_0 \qquad (x^\pm\equiv x^1\pm i x^2)\,. 
\end{eqnarray}
Then the anti-commutation relations containing $S$ are 
\begin{eqnarray}
&& \{Q_1^{0},S^*\}=-G_- \ , \quad 
\{Q_1^{m*},S\}=\{Q_1^{m},S\}=\{Q_1^{0},S\}=0 \ , \nonumber \\ 
&& \{Q_2^*,S\}=
-\frac{1}{2}D -\frac{i}{2}J
+\frac{3}{4}iR \ , \qquad \{S,S^*\}=K \ ,
\end{eqnarray}
where $R$ is an R-symmetry generator defined as 
\begin{eqnarray}
R\equiv -\int \! d^2x\, 
\mathrm{Tr}\left(\frac{2}{3}\phi^{\dagger}_a \phi^a 
- \frac{2}{3}\phi^{\dagger}_{a'} \phi^{a'} 
-\frac{1}{3}\psi^{\dagger a} \psi_a 
+ \frac{1}{3}\psi^{\dagger a'} \psi_{a'} \right) \ .
\end{eqnarray}
In fact, $R$ generates the $U(1)$ R-symmetry
\begin{eqnarray}
&& i[R,Q_1^0]= -iQ_1^0~,~~~
i[R,Q_1^m]=i\frac{1}{3}Q_1^m~, 
\nonumber \\ && 
i[R,Q_2]=-iQ_2~, \quad i[R,S]=  -iS~, 
\end{eqnarray}
and commutes with all bosonic generators
\begin{eqnarray}
[R,T_B]=0~,~~~
T_B=\{
P_i,~
H,~
J,~
G_i,~
D,~
K,~
M,~
R^{mn},~
R
\}~.
\end{eqnarray}
Finally the remaining non-trivial commutation relations are 
\begin{eqnarray}
&& i[P_-,S]= - Q_1^0 \ , \quad 
i[H,{S}]=-{Q_2} \ , \quad i[J,S]=-\frac{i}{2} S \ , \nonumber \\ 
&&  i[D,S]= S \ , \quad [G_i,S]=[K,S]=[M,S]=[R^{mn},S]=0 \ .
\end{eqnarray}

\subsection{Summary of the superconformal algebra}

We summarize the superconformal algebra with 14 fermionic generators
obtained in this section.
The bosonic part is nothing but the Schr\"odinger algebra:
\begin{align}
i[J,P_+] &= -i P_+ \ , \ \ i[J,P_-] = i P_- \ , \ \ i[J,G_+] = - iG_+ \ , \ \ i[J,G_-] = iG_- \ ,\cr
i[H,G_+] &= P_+ \ , \ \ i[H,G_-] = P_- \ , \ \ i[K,P_+] = - G_+ \ ,  \ \ i[K,P_-] = - G_- \ ,\cr
i[D,P_+] &= - P_+ \ , \ \ i[D,P_-] = - P_- \ , \ \ i[D,G_+] = G_+ \ , \ \ i[D,G_-] = G_- \ ,\cr
 i[D,H] &= -2H \ , \ \ i[H,K] = D \ , \ \ i[D,K] = 2K \ , \ \ i[P_+,G_-] = 2M \ .
\end{align}
The fermionic part is 
\begin{align}
\{Q^0_1 , Q_1^{0*}\} &= 2  M \ , \ \  \{Q_1^{m *}, Q_1^{n} \} = 2 M\delta^{mn}- 2mi R^{mn} \ , \cr 
\{Q_2,Q_2^*\} &= H \ , \ \ \{Q_1,Q_2^*\} = P_- \ , \ \ 
\{Q_2,{Q}_1^*\} = P_+ \ , \ \cr 
\ i[J,Q_1^{0}] &= \frac{i}{2}Q_1^0 \ , \ \ 
i[J,Q_1^{m}] = \frac{i}{2}Q_1^m \ , \ \ i[J,Q_2] = -\frac{i}{2}Q_2 \ , \cr
i[G_-,Q_2] &= -Q_1^{0} \ , \ \ i[G_+,Q_2^*] = -Q_1^{* 0} \ , \ \ i[D,Q_2] = -Q_2 \ , \ \ i[D,Q_2^*] = -Q_2^* \ , \cr
i[K,Q_2] &= S \ , \ \ i[H,S^*] = -Q_2^* \ , \ \ i[P_-,S] = -Q_1^{0} \ , \ \ i [J, S] = -\frac{i}{2} S \ , \cr
\{S,S^*\} &= K \ , \ \ \{S,Q_1^{*0} \} = -G_+ \ , \ \ i[D,S] = S \ , \ \ \{S,Q_2^*\} = \frac{i}{2}(iD - J +\frac{3}{2}R) \ , \cr
i[R,Q_1^0] &=  -i Q_1^0 \ , \ i[R,Q_1^m]= \frac{i}{3}Q_1^m \ , \ \ i[R,Q_2] = -iQ_2 \ , \ \  i[R,S] = -iS \ .
\end{align}

\section{Less SUSY limit}

In this section, we study other non-relativistic limits of the
mass deformed ABJM model, which lead to less supersymmetric theories. The
result is summarized in Table \ref{tab:1}. We only consider the
non-relativistic limit which preserves $SU(2) \times SU(2)$ global
symmetry while it is possible to obtain less and less SUSY limit by
breaking $SU(2) \times SU(2)$ global symmetry.

\begin{table}[tb]
\begin{center}
\begin{tabular}{c|c c c c|c|c|c}
 Limit  & $X^a $& $X^{a'}$ & $\Psi_a$ & $\Psi_{a'}$ & $Q_1$ &$Q_2$ &$S$ \\
 \hline
 $3$&  P        & P & P & P & $10$ &$2$& $2$  \\
  $4.1$& P  & A & A & P & $8$ &$0$ & $0$  \\
  $4.2$&  P & A & P & A & $4$ &$0$ & $0$ \\ 
$4.3$&  P & P & A & A & $0$ &$0$ & $0$ \\ 
 \end{tabular}
\end{center}
\caption{The matter contents of possible non-relativistic limits that
preserve $SU(2) \times SU(2)$ and non-trivial supersymmetries. P and A denote
particle and anti-particle, respectively.}  \label{tab:1}
\end{table}

\subsection{8 SUSY limit}

Let us take the ansatz for the non-relativistic limit of scalars as
\begin{align}
X^a = \frac{1}{\sqrt{2m}} e^{-imt} \phi^a \ , \qquad 
X^{a'} = \frac{1}{\sqrt{2m}} e^{imt} \hat{\phi}^{*a'} \ , 
\end{align}
and fermions as
\begin{align}
\Psi_a = e^{imt} \sigma_2\hat{\psi}_a^* \ ,  \qquad 
\Psi_{a'} = e^{-imt} \psi_{a'} \ .
\end{align}
The Dirac equation for $\Psi_{a}$ gives slightly different results from
those in section 3:
\begin{eqnarray}
\Psi_{a} = e^{imt} \left( 
  \begin{array}{c}
   -i\hat{\psi}^{*}_a \\ 
    i\frac{D_-}{2m} \hat{\psi}^*_a    
  \end{array}
\right) \ .
\end{eqnarray}
The action is given by $S_{\rm CS} + S_{\rm kin} + S_{\rm bos} 
+ S_{\rm fer}$, where
$S_{\rm CS}$ is the same as in \eqref{CSa} while the kinetic term is given
by
\begin{align}
S_{\rm kin} &= \int \!dt d^2x\, \left[i\mathrm{Tr} (\phi^\dagger_a D_0\phi^a 
+ \hat{\phi}^{\dagger a'} D_0 \hat{\phi}_{a'}) 
- \frac{1}{2m} \mathrm{Tr} (D_i\phi^\dagger_a D_i \phi^a 
+ D_i \hat{\phi}^{\dagger a'} D_i \hat{\phi}_{a'}) \right. \cr
& \qquad 
+ \left. i\mathrm{Tr} (\hat{\psi}^{\dagger}_a D_0 \hat{\psi}^a 
 + {\psi}^{\dagger a'} D_0 {\psi}_{a'}) 
+ \frac{1}{2m} \mathrm{Tr} (\hat{\psi}^{\dagger}_a D_- D_+ \hat{\psi}^a 
+ {\psi}^{\dagger a'} D_+D_- {\psi}_{a'}) \right] . \label{kine1}
\end{align}
Now, $\phi^a$ and $\psi_{a'}$ transform as $(N,\bar{N})$ under
$U(N)\times U(N)$ whereas $\hat{\phi}_{a'}$ and $\hat{\psi}^{a}$
transform as $(\bar{N},N)$ \ .

The leading bosonic potential that will survive in the conformal limit is 
\begin{eqnarray}
S_{\rm bos} = \int \!dt d^2x\, \frac{\pi}{km} 
\mathrm{Tr} (\phi^a\phi^\dagger_{[a} \phi^b \phi^\dagger_{b]} 
- \hat{\phi}^{\dagger a'}\hat{\phi}_{[a'} \hat{\phi}^{\dagger b'} 
\hat{\phi}_{b']}) \ . \label{bos1}
\end{eqnarray}

The fermionic potential comes from the non-relativistic limit of \eqref{ferp}:
\begin{align}
S_{\rm fer} &= - \frac{\pi }{km}\int \!dt d^2x\,  \mathrm{Tr} 
\left[ (\phi^\dagger_a\phi^a + \hat{\phi}_{a'} 
\hat{\phi}^{\dagger a'})(\hat{\psi}^b \hat{\psi}^{\dagger}_b 
+  {\psi}^{\dagger b'} {\psi}_{b'} )  \right. \cr 
& \qquad  +(\phi^a\phi^\dagger_a 
+\hat{\phi}^{\dagger a'}\hat{\phi}_{a'}
 )(\hat{\psi}^{\dagger}_b\hat{\psi}^b 
+  {\psi}_{b'} {\psi}^{\dagger b'}) \cr
& \qquad - 2 (\phi^a\phi^\dagger_b \hat{\psi}^\dagger_a\hat{\psi}^{b} 
-i \phi^a \hat{\phi}_{b'} \hat{\psi}^\dagger_a \psi^{\dagger b'} 
+ i \hat{\phi}^{\dagger a'} \phi_b^\dagger \psi_{a'} \hat{\psi}^b 
+ \hat{\phi}^{\dagger a'} \hat{\phi}_{b'} {\psi}_{a'}{\psi}^{\dagger b'}) \cr
& \qquad \left. - 2(\phi^\dagger_a \phi^b \hat{\psi}^a \hat{\psi}_b^\dagger 
+ i \phi_a^\dagger \hat{\phi}^{\dagger b'} \hat{\psi}^a \psi_{b'} 
-i \hat{\phi}_{a'} \phi^b \psi^{\dagger a'} \hat{\psi}^\dagger_b 
+ \hat{\phi}_{a'} \hat{\phi}^{\dagger b'} \psi^{\dagger a'} \psi_b)\right] \ .
\end{align}

Let us study the bosonic symmetry of the theory. The theory possesses
the full Schr\"odinger symmetry and $SU(2) \times SU(2)$ R-symmetry
acting on indices $a$ and $a'$. In addition, the theory is invariant
under $U(1)_B$ and $U(1)_F$ generated by $Q_B (\phi^a, \hat{\phi}_{a'},
\hat{\psi}^a, {\psi}_{a'}) = (1,-1,0,0)$ and $Q_F(\phi^a,
\hat{\phi}_{a'}, \hat{\psi}^a, {\psi}_{a'}) = (0,0,1,-1)$. Furthermore,
because $\epsilon^{ab}$ and $\epsilon^{a'b'}$ do not appear in the
action, the $SU(2)\times SU(2)$ symmetry is enhanced to $U(2) \times
U(2)$ with additional $U(1)_R$ charge generated by $Q_{R_1}(\phi^a,
\hat{\phi}_{a'}, \hat{\psi}^a, {\psi}_{a'}) = (1,0,1,0)$ and $Q_{R_2}
(\phi^a, \hat{\phi}_{a'}, \hat{\psi}^a, {\psi}_{a'}) =
(0,1,0,1)$.\footnote{Since there are two relations: $M= Q_{R_1}-Q_{R_2}$ and
$Q_B +Q_F= Q_{R_1}+Q_{R_2}$, the total symmetry is $U(2) \times U(2) \times
U(1)_F$. In addition, a particular $U(1) \times U(1)$ is a part of the
gauge symmetry.}

We now consider the SUSY transformation. The supersymmetries generated by
$\Gamma^{1}$ and $\Gamma^2$ do not act on the fields non-trivially any
longer because the particles cannot transform into anti-particles in the
non-relativistic limit. The only non-trivial SUSY transformations are
generated by $\Gamma^{3-6}$.

The corresponding SUSY generators are
\begin{align}
Q^{3}_1 &= \sqrt{2m}i \int \!d^2x\, \mathrm{Tr}\left(-i\phi^1 \psi^{\dagger2'} + i\phi^2 \psi^{\dagger 1'} - \hat{\phi}^{\dagger 1'} \hat{\psi}^{\dagger 2}+ \hat{\phi}^{\dagger 2'} \hat{\psi}^{\dagger 1} \right) \ , \cr
Q^{4}_1 &= \sqrt{2m} i \int \!d^2x\, \mathrm{Tr} \left(\phi^1 \psi^{\dagger 2'} + \phi^{2} \psi^{\dagger 1'} +i\hat{\phi}^{\dagger 1'} \hat{\psi}^2 +i \hat{\phi}^{\dagger 2'} \hat{\psi}^1 \right) \ , \cr
Q^{5}_1 &= \sqrt{2m} i \int\! d^2x\, \mathrm{Tr} \left( -\phi^1 \psi^{\dagger 1'} + \phi^2 \psi^{\dagger 2'} -i  \hat{\phi}^{\dagger 1'} \hat{\psi}^1 +i\hat{\phi}^{\dagger 2'} \hat{\psi}^2 \right) \ ,  \cr
Q^{6}_1 &= \sqrt{2m} i \int \!d^2x\, \mathrm{Tr} \left(i \phi^1 \psi^{\dagger 1'} +i\phi^2 \psi^{\dagger 2'} - \hat{\phi}^{\dagger 1'} \hat{\psi}^1 -\hat{\phi}^{\dagger 2'} \hat{\psi}^2 \right) \ .
\end{align}

We can compute the anti-commutation relations as
\begin{align}
\{Q_1^{m*}, Q_1^{n} \} &= 2M\delta^{mn} - 2mi R^{mn} \ , \qquad 
i[J, Q_1^m] = \frac{i}{2} Q_1^m  \ , \cr
[H,Q_1^{n}] &= [P_i,Q_1^{n}] = [G_i,Q_1^{n}] = [D,Q_1^{n}] = [K,Q_1^{n}] = [M,Q_1^{n}] = 0 \ .
\end{align}
$R^{mn}$ are particular combinations of the $SU(2)\times SU(2)$ R-charges:
\begin{align}
R^{34} &= \int d^2 x \mathrm{Tr}(
- \psi^{\dagger 2'} \psi_{2'} 
+ \psi^{\dagger 1'}\psi_{1'} 
- \hat{\psi}^2 \hat{\psi}^{\dagger}_{2} 
+ \hat{\psi}^1 \hat \psi^{\dagger}_1 
- \phi^1 \phi_1^\dagger 
+ \phi^2 \phi_2^\dagger 
+ \hat{\phi}^{\dagger 1'} \hat{\phi}_{1'} 
- \hat{\phi}^{\dagger 2'} \hat{\phi}_{2'})\,, \cr
R^{35} &= \int d^2 x \mathrm{Tr}(\psi^{\dagger 2'} \psi_{1'} + \psi^{\dagger 1'}\psi_{2'} + \hat{\psi}^2 \hat{\psi}^{\dagger}_{1} + \hat{\psi}^1 \hat{\psi}^{\dagger}_{2} - \phi^1 \phi_2^\dagger - \phi^2 \phi_1^\dagger + \hat{\phi}^{\dagger 1'} \hat{\phi}_{2'}+ \hat{\phi}^{\dagger 2'} \hat{\phi}_{1'})\,, \cr
 R^{36} &= \int d^2 x \mathrm{Tr}(i \psi^{\dagger 2'} \psi_{1'} - i\psi^{\dagger 1'}\psi_{2'} +i \hat{\psi}^2 \hat{\psi}^{\dagger}_{1} -i \hat{\psi}^1 \hat{\psi}^{\dagger}_{2} +i \phi^1 \phi_2^\dagger -i \phi^2 \phi_1^\dagger -i \hat{\phi}^{\dagger 1'} \hat{\phi}_{2'} +i \hat{\phi}^{\dagger 2'} \hat{\phi}_{1'})\,, \cr
 R^{45} &= \int d^2 x \mathrm{Tr}(i \psi^{\dagger 2'} \psi_{1'} - i\psi^{\dagger 1'}\psi_{2'} -i \hat{\psi}^2 \hat{\psi}^{\dagger}_{1} +i \hat{\psi}^1 \hat{\psi}^{\dagger}_{2} -i \phi^1 \phi_2^\dagger +i \phi^2 \phi_1^\dagger -i \hat{\phi}^{\dagger 1'} \hat{\phi}_{2'} +i \hat{\phi}^{\dagger 2'} \hat{\phi}_{1'})\,, \cr
R^{46} &= \int d^2 x \mathrm{Tr}(-\psi^{\dagger 2'} \psi_{1'} - \psi^{\dagger 1'}\psi_{2'} + \hat{\psi}^2 \hat{\psi}^{\dagger}_{1} + \hat{\psi}^1 \hat{\psi}^{\dagger}_{2} - \phi^1 \phi_2^\dagger - \phi^2 \phi_1^\dagger - \hat{\phi}^{\dagger 1'} \hat{\phi}_{2'} - \hat{\phi}^{\dagger 2'} \hat{\phi}_{1'})\,, \cr
R^{56} &= \int d^2 x \mathrm{Tr}(
-\psi^{\dagger 2'} \psi_{2'} 
+ \psi^{\dagger 1'}\psi_{1'} 
+ \hat{\psi}^2 \hat{\psi}^{\dagger}_{2} 
- \hat{\psi}^1 \hat{\psi}^{\dagger}_{1} 
{+} \phi^1 \phi_1^\dagger 
{-} \phi^2 \phi_2^\dagger 
{+} \hat{\phi}^{\dagger 1'} \hat{\phi}_{1'} 
{-} \hat{\phi}^{\dagger 2'} \hat{\phi}_{2'}) \,.
\end{align}

We cannot construct a dynamical SUSY charge $Q_2$ and hence there is
no superconformal generator $S$. This gives us an example of
non-relativistic superconformal field theories with no superconformal
charges.

\subsection{4 SUSY limit}

We take the ansatz for the non-relativistic limit of scalars as
\begin{align}
X^a = \frac{1}{\sqrt{2m}} e^{-imt} \phi^a \ , \qquad 
X^{a'} = \frac{1}{\sqrt{2m}} e^{imt} \hat{\phi}^{*a'} \ , 
\end{align}
and fermions as
\begin{align}
\Psi_a = e^{-imt} \psi_a \ , \qquad 
\Psi_{a'} = e^{imt} \sigma_2 \hat{\psi}_{a'}^* \ .
\end{align}
The Dirac equation for $\Psi_{a'}$ gives slightly different results from
those in section 3:
\begin{eqnarray}
\Psi_{a'} = e^{imt} \left( 
  \begin{array}{c}
   \frac{iD_+}{2m} \hat{\psi}_{a'}^* \\ 
    i\hat{\psi}_{a'}^*   
  \end{array}
\right) \ .
\end{eqnarray}
The action is given by $S_{\rm CS} + S_{\rm kin} + S_{\rm bos} + S_{\rm fer}$, where
$S_{\rm CS}$ is the same as in \eqref{CSa} while the kinetic term is given
by
\begin{align}
S_{\rm kin} &= \int\! dt d^2x\, \left[i\mathrm{Tr} (\phi^\dagger_a D_0 \phi^a + \hat{\phi}^{\dagger a'} D_0 \hat{\phi}_{a'}) - \frac{1}{2m} \mathrm{Tr} (D_i\phi^\dagger_a D_i \phi^a + D_i \hat{\phi}^{\dagger a'} D_i \hat{\phi}_{a'}) \right. \cr
& \qquad + \left. i\mathrm{Tr} (\psi^{\dagger a} D_0 \psi_a + \hat{\psi}^\dagger_{a'} D_0 \hat{\psi}^{a'}) + \frac{1}{2m} \mathrm{Tr} (\psi^{\dagger a} D_- D_+ \psi_a + \hat{\psi}^\dagger_{a'} D_+D_- \hat{\psi}^{a'}) \right] . \label{kine2}
\end{align}
Now, $\phi^a$ and $\psi_a$ transform as $(N,\bar{N})$ under $U(N)\times
U(N)$ whereas $\hat{\phi}_{a'}$ and $\hat{\psi}^{a'}$ transform as
$(\bar{N},N)$. This is equivalent to the non-relativistic limit studied
in \cite{Nakayama:2008qm}.

The leading bosonic potential that will survive in the conformal limit is
\begin{eqnarray}
S_{\rm bos} = \frac{\pi}{km} \int\! dt d^2x\,  \mathrm{Tr} (\phi^a\phi^\dagger_{[a} \phi^b \phi^\dagger_{b]} - \hat{\phi}^{\dagger a'}\hat{\phi}_{[a'} \hat{\phi}^{\dagger b'} \hat{\phi}_{b']}) \ . \label{bosa2}
\end{eqnarray}
The fermionic potential comes from the non-relativistic limit of \eqref{ferp}:
\begin{align}
S_{\rm fer} &= \frac{\pi }{km}\int \!dt d^2x\,   \mathrm{Tr} \left[ (\phi^\dagger_a\phi^a + \hat{\phi}_{a'} \hat{\phi}^{\dagger a'})(\psi^{\dagger b}\psi_b +  \hat{\psi}^{b'} \hat{\psi}^\dagger_{b'})   \right.  \cr 
& \qquad +(\phi^a\phi^\dagger_a +\hat{\phi}^{\dagger a'}\hat{\phi}_{a'} )(\psi_b \psi^{\dagger b} +  \hat{\psi}^\dagger_{b'} \hat{\psi}^{b'} )\cr
&\qquad \left .- 2 (\phi^a\phi^\dagger_b \psi_a\psi^{\dagger b} + \hat{\phi}^{\dagger a'} \hat{\phi}_{b'} \hat{\psi}^\dagger_{a'} \hat{\psi}^{ b'}) - 2(\phi^\dagger_a \phi^b\psi^{\dagger a}\psi_b +\hat{\phi}_{a'} \hat{\phi}^{\dagger b'} \hat{\psi}^{a'} \hat{\psi}^\dagger_{b'}) \right] \ .
\end{align}

Let us study the bosonic symmetry of the theory. The theory possesses
the full Schr\"odinger symmetry and $SU(2) \times SU(2)$ R-symmetry
acting on indices $a$ and $a'$. In addition, the theory is invariant
under $U(1)_B$ and $U(1)_F$ generated by $Q_B (\phi^a, \hat{\phi}_{a'},
\psi_a, \hat{\psi}^{a'}) = (1,-1,0,0)$ and $Q_F(\phi^a, \hat{\phi}_{a'},
\psi_a, \hat{\psi}^{a'}) = (0,0,1,-1)$. Furthermore, because
$\epsilon^{ab}$ and $\epsilon^{a'b'}$ do not appear in the action, the
$SU(2)\times SU(2)$ symmetry is enhanced to $U(2) \times U(2)$ with
additional $U(1)_R$ charge generated by $Q_{R_1}(\phi^a,
\hat{\phi}_{a'}, \psi_a, \hat{\psi}^{a'}) = (1,0,-1,0)$ and
$Q_{R_2}(\phi^a, \hat{\phi}_{a'}, \psi_a, \hat{\psi}^{a'}) =
(0,1,0,-1)$.\footnote{There is one relation between $U(1)$ charges:
$Q_{R_1} - Q_{R_{2}} = Q_B -Q_F$, so the total symmetry is $U(2) \times
U(2) \times U(1)_F \times U(1)_M$. In other words, the $U(1)$ symmetries
are generated by all the independent rotations of $(\phi^a,
\hat{\phi}_{a'}, \psi_a, \hat{\psi} ^{a'})$. A particular combination of
$U(1) \times U(1)$ is a part of the gauge symmetry.}
 
While the bosonic sector has a larger symmetry than the limit discussed
in section 3, the number of supersymmetry is reduced. This is due to the fact that
the supersymmetries generated by $\Gamma^{3-6}$ do not act on the fields
non-trivially any longer because the particle cannot transform into
anti-particle in the non-relativistic limit. The only non-trivial SUSY
transformations are generated by $\Gamma^1$ and $\Gamma^2$.  The kinematical SUSY charges
are
\begin{align}
Q_1 &(\equiv Q_1^1+iQ_1^2)= \sqrt{2m}i  \int \!d^2x\, \mathrm{Tr} \left(i \phi_1^\dagger \psi_2 - i\phi^\dagger_2 \psi_1 \right) 
\ , \cr
\hat Q_1 &(\equiv Q_1^1-iQ_1^2)= \sqrt{2m}i  \int\! d^2x\, \mathrm{Tr} \left(- \hat{\phi}_{1'} \hat{\psi}^\dagger_{2'} +\hat{\phi}_{2'} \hat{\psi}^\dagger_{1'} \right)
\ ,
\end{align}
and there is no dynamical SUSY. As a consequence, there is no superconformal symmetry. The anti-commutation relations are
\begin{align}
\{Q_1^{*},Q_1 \} = 2M_P \ , \ \ \{\hat Q_1^{*},\hat Q_1 \} =2 \hat{M}_A \ , \ \
\{Q_1,\hat Q_1\} = \{Q_1^{*},\hat Q_1\} = 0 \ ,
\end{align}
where $M_P$ is the mass operator for particles, and $\hat{M}_A$ is the mass
operator for anti-particles.

\subsection{0 SUSY limit}

We can construct a non-supersymmetric theory by taking the
non-relativistic ansatz
\begin{eqnarray}
X^A = \frac{1}{\sqrt{2m}}e^{-imt} \phi^A 
\end{eqnarray}
and
\begin{eqnarray}
\Psi_A = e^{imt} \sigma_2 \hat{\psi}^*_A \ .
\end{eqnarray}
It is clear that since the bosons are all particles and fermions are all
anti-particles, there is no non-trivial supersymmetry acting on the
non-relativistic theory.

Without writing down the action explicitly, we just point out that the
bosonic symmetry is given by the Schr\"odinger algebra with global
$SU(2)\times SU(2) \times U(1)_B \times U(1)_F$ symmetries. Due to the
lack of the supersymmetry, however, it is quite probable that the model
breaks the conformal invariance at the quantum level. Conformal invariance of the non-relativistic Chern-Simons-Matter theory has been discussed in \cite{Bergman:1993kq,Kim:1996rz,deKok:2008ge}.

\section{Discussion and Summary}

In this paper, we have studied various non-relativistic limits of the
$\mathcal{N}=6$ superconformal field theories and constructed different
non-relativistic conformal field theories. While the kinematical SUSY is easy to obtain, the emergence of the dynamical SUSY is non-trivial. We need a specific combination of the
relativistic supersymmetry whose leading order supersymmetry
transformation vanishes in the non-relativistic limit.

One may try to obtain more supersymmetries by starting with Bagger-Lambert
$\mathcal{N}=8$ supersymmetric Chern-Simons theory
\cite{Bagger:2007jr,Bagger:2007vi}. Again it is not so difficult
to construct the limit where only the kinematical SUSY remains
while it is still an open question whether we could obtain more
dynamical supersymmetries there.

Given a new non-relativistic superconformal algebra, one could define
a (non-relativistic) superconformal index \cite{Nakayama:2008qm}, and
compute it from the explicit theory we have constructed in this
paper. The superconformal algebras we have obtained in this paper have a
non-trivial involutive anti-automorphism, so it is straight-forward to
define a new class of indices.

Finally, the supergravity dual of the non-relativistic limit of the ABJM
theory is of most importance for a future study. The existence of
several different non-relativistic limits, as we have discussed in this
paper, suggests that corresponding different non-relativistic limits
should also exist in the dual supergravity solution. It would be very
interesting to pursue this direction further. Some related supergravity
backgrounds with Schr\"odinger (super)symmetry have been studied in
\cite{Son:2008ye,Balasubramanian:2008dm,Goldberger:2008vg,Barbon:2008bg,
Wen:2008hi,Herzog:2008wg,Maldacena:2008wh,Adams:2008wt,KLM,HY,
Schvellinger,Mazzucato,RRST,AMSV,Taylor,G,Alishahiha:2009hg,AADV}.

\section*{Acknowledgments}

We would like to thank S. Ryu for fruitful discussions. The work of NY
was supported in part by the National Science Foundation under Grant
No. PHY05-55662 and the UC Berkeley Center for Theoretical Physics. The
work of MS was supported in part by the Grant-in-Aid for Scientific
Research (19540324) from the Ministry of Education, Science and Culture,
Japan. The work of KY was supported in part by the National Science
Foundation under Grant No. PHY05-51164 and JSPS Postdoctoral Fellowships
for Research Abroad.  

\appendix

\section{Spinor convention}

We take the $(-,+,+)$ metric convention and chiral representation of the
gamma matrix in $(1+2)$ dimension: $\gamma^\mu = (i\sigma_3, \sigma_1,
-\sigma_2)$. They satisfy the Clifford algebra $\{\gamma^\mu,
\gamma^\nu\} = 2\eta^{\mu\nu}$.  The Dirac conjugation is given by
$\bar{\psi} = \psi^\dagger \gamma^0 = \psi^\dagger i \sigma^3$. The
corresponding scalar product is $\bar{\psi} \psi \equiv i\psi^*_\alpha
\sigma_3^{\alpha \beta} \psi_\beta $. We can define a raised spinor by
$\psi^\alpha = \epsilon^{\alpha\beta} \psi_\beta =
i\sigma_2^{\alpha\beta} \psi_\beta$ so that $\chi \psi \equiv
\chi^{\alpha} \psi_{\alpha}$ is a Lorentz scalar.  Similarly we define
$\psi^\dagger \chi^\dagger \equiv \epsilon^{\alpha \beta}
\psi^\dagger_\beta \chi^\dagger_\alpha = -(\chi \psi)^*$.

In this chiral basis, the Majorana condition is imposed by $\alpha
\sigma_1 \psi^* = \psi$ with $|\alpha|^2 = 1$. We choose $\alpha =-i$
with no loss of generality.

\section{Consistency of the truncation}

In this appendix, we address the consistency of the non-relativistic
truncation studied in the main text. When we substitute the
non-relativistic ansatz with both particles and anti-particles into the
relativistic action, we have non-trivial interactions that might
induce inconsistency.

Having Non-Abelian ordering of the operators and index structure
suppressed, which are irrelevant for this study, we have the following
interactions in the relativistic theory
\begin{align}
& (\phi_a \phi_{b'}^* + \hat{\phi}^*_c\hat{\phi}_{d'})(\psi_e
 \psi_{f'}^* + \hat{\psi}_g^* \hat{\psi}_{h'}) + {\rm c.c.} \ , \cr
&\phi^*_a \hat{\phi}_b \psi_c \hat{\psi}_d^* + {\rm c.c.} \ , \quad  
\phi^*_{a'} \hat{\phi}_{b'} \psi_{c'} \hat{\psi}_{d'}^* + {\rm c.c.} \ , \cr
& \phi^*_{a'} \psi_{b'} \hat{\phi}_{c} \hat{\psi}_d^* + {\rm c.c.} \ , \quad 
\phi^*_{a'} \psi_{b} \hat{\phi}_{c'} \hat{\psi}_d^* + {\rm c.c.} \ , \cr
 & \phi^*_{a'} \psi_{b} \hat{\phi}_{c} \hat{\psi}_{d'}^* + {\rm c.c.} \ , \quad 
\phi^*_{a} \psi_{b'} \hat{\phi}_{c'} \hat{\psi}_d^* + {\rm c.c.} \ , \cr
& \phi^*_{a} \psi_{b'} \hat{\phi}_{c} \hat{\psi}_{d'}^* + {\rm c.c.} \ , \quad 
\phi^*_{a} \psi_{b} \hat{\phi}_{c'} \hat{\psi}_{d'}^* + {\rm c.c.}  \ .
\end{align}

As discussed in \cite{Nakayama:2008qz}, we can impose either the strong condition,
which means the conservation of the particle number, or the weak
condition, which means the consistency at the level of classical
equation of motion. The former is strong because there could be no
quantum creation of particles, but the latter truncation is still
consistent as a classical theory because it does not provide any source
for discarded fields.

We see that the PPPP truncation (section 3) is consistent under the
strong condition while PAAP (section 4.1), PAPA (section 4.2) and PPAA
(section 4.3) truncations are only consistent under the weak
condition. We could imagine the truncation which does not satisfy any
condition such as PPPA truncation. While there is no problem in finding
classical Schr\"odinger invariant field theories from such a
construction, the supersymmetry is typically broken.


\begin{thebibliography}{99}

\bibitem{Gustavsson:2007vu}
  A.~Gustavsson,
  arXiv:0709.1260 [hep-th].

\bibitem{Bagger:2007jr}
  J.~Bagger and N.~Lambert,
  Phys.\ Rev.\  D {\bf 77}, 065008 (2008)
  [arXiv:0711.0955 [hep-th]].

\bibitem{Bagger:2007vi}
  J.~Bagger and N.~Lambert,
  JHEP {\bf 0802}, 105 (2008)
  [arXiv:0712.3738 [hep-th]].

\bibitem{Aharony:2008ug}
  O.~Aharony, O.~Bergman, D.~L.~Jafferis and J.~Maldacena,
  JHEP {\bf 0810}, 091 (2008)
  [arXiv:0806.1218 [hep-th]].

\bibitem{Fujita}
  M.~Fujita, W.~Li, S.~Ryu and T.~Takayanagi,
  arXiv:0901.0924 [hep-th].

\bibitem{Aharony:2008gk}
  O.~Aharony, O.~Bergman and D.~L.~Jafferis,
  arXiv:0807.4924 [hep-th].

\bibitem{Giveon:2008zn}
  A.~Giveon and D.~Kutasov,
  arXiv:0808.0360 [hep-th].
  
\bibitem{Sakaguchi:2008rx}
  M.~Sakaguchi and K.~Yoshida,
  J.\ Math.\ Phys.\  {\bf 49} (2008) 102302
  [arXiv:0805.2661 [hep-th]].

\bibitem{Sakaguchi:2008ku}
  M.~Sakaguchi and K.~Yoshida,
  JHEP {\bf 0808}, 049 (2008)
  [arXiv:0806.3612 [hep-th]].

\bibitem{Hagen:1972pd}
  C.~R.~Hagen,
  Phys.\ Rev.\  D {\bf 5}, 377 (1972).
  
\bibitem{Niederer:1972zz}
  U.~Niederer,
  Helv.\ Phys.\ Acta {\bf 45} (1972) 802.

\bibitem{Perroud:1977qh}
  M.~Perroud,
  Helv.\ Phys.\ Acta {\bf 50} (1977) 233.
  
\bibitem{Barut:1981mt}
  A.~O.~Barut and B.~W.~Xu,
  Phys.\ Lett.\  A {\bf 82} (1981) 218.

\bibitem{Hussin:1986cc}
  V.~Hussin and M.~Jacques,
  J.\ Phys.\ A  {\bf 19}, 3471 (1986).
  
\bibitem{Jackiw:1990mb}
  R.~Jackiw and S.~Y.~Pi,
  Phys.\ Rev.\  D {\bf 42}, 3500 (1990)
  [Erratum-ibid.\  D {\bf 48}, 3929 (1993)].
  
\bibitem{Jackiw:1992fg}
  R.~Jackiw and S.~Y.~Pi,
  arXiv:hep-th/9206092.
  
\bibitem{Henkel:1993sg}
  M.~Henkel,
  J.\ Statist.\ Phys.\  {\bf 75}, 1023 (1994)
  [arXiv:hep-th/9310081].


\bibitem{Dobrev}
V.~K.~Dobrev, H.~D.~Doebner and C.~Mrugalla,
Reports on mathematical physics, 39, 201 (1997)
\bibitem{Mehen:1999nd}
  T.~Mehen, I.~W.~Stewart and M.~B.~Wise,
  Phys.\ Lett.\  B {\bf 474}, 145 (2000)
  [arXiv:hep-th/9910025].

\bibitem{Nishida:2007pj}
  Y.~Nishida and D.~T.~Son,
  Phys.\ Rev.\  D {\bf 76}, 086004 (2007)
  [arXiv:0706.3746 [hep-th]].

\bibitem{Leblanc:1992wu}
  M.~Leblanc, G.~Lozano and H.~Min,
  Annals Phys.\  {\bf 219}, 328 (1992)
  [arXiv:hep-th/9206039].

\bibitem{Nakayama:2008qm}
  Y.~Nakayama,
  JHEP {\bf 0810}, 083 (2008)
  [arXiv:0807.3344 [hep-th]].
  
\bibitem{Nakayama:2008qz}
Y.~Nakayama, S.~Ryu, M.~Sakaguchi and K.~Yoshida,
  JHEP {\bf 0901} (2009) 006
  [arXiv:0811.2461 [hep-th]].

\bibitem{NRSY}
  Y.~Nakayama, M.~Sakaguchi and K.~Yoshida,
  arXiv:0812.1564 [hep-th].

\bibitem{Hosomichi:2008jd}
  K.~Hosomichi, K.~M.~Lee, S.~Lee, S.~Lee and J.~Park,
  JHEP {\bf 0807}, 091 (2008)
  [arXiv:0805.3662 [hep-th]].

\bibitem{Hosomichi:2008jb}
  K.~Hosomichi, K.~M.~Lee, S.~Lee, S.~Lee and J.~Park,
  JHEP {\bf 0809}, 002 (2008)
  [arXiv:0806.4977 [hep-th]].

\bibitem{Gomis:2008vc}
  J.~Gomis, D.~Rodriguez-Gomez, M.~Van Raamsdonk and H.~Verlinde,
  JHEP {\bf 0809}, 113 (2008)
  [arXiv:0807.1074 [hep-th]].

\bibitem{Gaiotto:2008cg}
  D.~Gaiotto, S.~Giombi and X.~Yin,
  arXiv:0806.4589 [hep-th].

\bibitem{Terashima:2008sy}
  S.~Terashima,
  JHEP {\bf 0808}, 080 (2008)
  [arXiv:0807.0197 [hep-th]].

\bibitem{Bandres:2008ry}
  M.~A.~Bandres, A.~E.~Lipstein and J.~H.~Schwarz,
  JHEP {\bf 0809}, 027 (2008)
  [arXiv:0807.0880 [hep-th]].

\bibitem{Bergman:1993kq}
  O.~Bergman and G.~Lozano,
  Annals Phys.\  {\bf 229}, 416 (1994)
  [arXiv:hep-th/9302116].

\bibitem{Kim:1996rz}
  S.~J.~Kim and C.~k.~Lee,
  Phys.\ Rev.\  D {\bf 55}, 2227 (1997)
  [arXiv:hep-th/9606054].
\bibitem{deKok:2008ge}
  M.~O.~de Kok and J.~W.~van Holten,
  Nucl.\ Phys.\  B {\bf 805}, 545 (2008)
  [arXiv:0806.3358 [hep-th]].

\bibitem{Son:2008ye}
  D.~T.~Son,
  arXiv:0804.3972 [hep-th].
  
\bibitem{Balasubramanian:2008dm}
  K.~Balasubramanian and J.~McGreevy,
  arXiv:0804.4053 [hep-th].
  
\bibitem{Goldberger:2008vg}
  W.~D.~Goldberger,
  arXiv:0806.2867 [hep-th].
  
\bibitem{Barbon:2008bg}
  J.~L.~B.~Barbon and C.~A.~Fuertes,
  arXiv:0806.3244 [hep-th].
  
\bibitem{Wen:2008hi}
  W.~Y.~Wen,
  arXiv:0807.0633 [hep-th].
  
\bibitem{Herzog:2008wg}
  C.~P.~Herzog, M.~Rangamani and S.~F.~Ross,
  arXiv:0807.1099 [hep-th].
  
\bibitem{Maldacena:2008wh}
  J.~Maldacena, D.~Martelli and Y.~Tachikawa,
  arXiv:0807.1100 [hep-th].
  
\bibitem{Adams:2008wt}
  A.~Adams, K.~Balasubramanian and J.~McGreevy,
  arXiv:0807.1111 [hep-th].

 \bibitem{KLM}
  S.~Kachru, X.~Liu and M.~Mulligan,
  arXiv:0808.1725 [hep-th].  
  
\bibitem{HY}
  S.~A.~Hartnoll and K.~Yoshida,
  arXiv:0810.0298 [hep-th]. 
  
\bibitem{Schvellinger}
  M.~Schvellinger,
  arXiv:0810.3011 [hep-th].  

\bibitem{Mazzucato}
  L.~Mazzucato, Y.~Oz and S.~Theisen,
  arXiv:0810.3673 [hep-th].
  
\bibitem{RRST}
  M.~Rangamani, S.~F.~Ross, D.~T.~Son and E.~G.~Thompson,
  JHEP {\bf 0901} (2009) 075
  [arXiv:0811.2049 [hep-th]].   
  
\bibitem{AMSV}
  A.~Adams, A.~Maloney, A.~Sinha and S.~E.~Vazquez,
  arXiv:0812.0166 [hep-th].  
  
\bibitem{Taylor}
  M.~Taylor,
  arXiv:0812.0530 [hep-th].  
  
\bibitem{G}
  A.~Donos and J.~P.~Gauntlett,
  arXiv:0901.0818 [hep-th].  
\bibitem{Alishahiha:2009hg}
  M.~Alishahiha and A.~Ghodsi,
  arXiv:0901.3431 [hep-th].

\bibitem{AADV}
  A.~Akhavan, M.~Alishahiha, A.~Davody and A.~Vahedi,
  arXiv:0902.0276 [hep-th].  
  
\end{thebibliography}
\end{document}